\documentclass[aps,prb,twocolumn]{revtex4}

\usepackage{amsmath}
\usepackage{amssymb}
\usepackage{graphicx}
\usepackage{multirow}
\usepackage{array}

\begin{document}
    \author{X. Fabr\`eges$^1$, I. Mirebeau$^1$, S. Petit$^1$, P. Bonville$^2$,
     A. A. Belik$^3$ }
    \date{5/04/2011}
    \affiliation{$^1$ CEA, Centre de Saclay, /DSM/IRAMIS/ Laboratoire L\'eon Brillouin,
     91191 Gif-sur-Yvette,France}
    \affiliation{$^2$ CEA, Centre de Saclay, /DSM/IRAMIS/ Service de Physique de l'Etat Condens\'e, 91191 Gif-Sur-Yvette, France}
     \affiliation{$^3$ International Center for Materials Nanoarchitectonics (MANA), National Institute for Material Science (NIMS), 1-1 Namiki, Tsukuba , Ibaraki 305-0044 Japan}
    \title{InMnO$_3$ : a fully frustrated multiferroic.}

    \begin{abstract}
    InMnO$_3$ is a peculiar member of the hexagonal manganites h-RMnO$_3$ (where R is a rare earth metal element), showing crystalline, electronic and magnetic properties at  variance with the other compounds of the family. We have studied high quality samples synthesized at high pressure and temperature by powder neutron diffraction.  The position of the Mn ions is found to be close to the threshold $\it{x}=1/3$ where superexchange Mn-Mn interactions along the $\it{c}$ axis compensate. Magnetic long range order occurs below $T_{\rm N}$= 120(2)~K with a magnetic unit cell doubled along $\it{c}$, whereas short range two dimensional dynamical spin correlations are observed above $T_{\rm N}$. We propose that pseudo-dipolar interactions are responsible for the long period magnetic structure.
    \end{abstract}

    \maketitle

    \section{Introduction}
		Multiferroic systems have been intensively studied in the past ten years as the coupling between ferroelectric and magnetic order parameters may lead to novel electronic devices. This coupling can have different microscopic origins, related either to Dzyaloshinskii-Moriya interactions\cite{Katsura2005, Sergienko2006} or to an exchange-striction mechanism,  and it is still not fully understood. All multiferroics show complex and mostly non collinear magnetic orders, arising from competing interactions and/or geometrical frustration.

The hexagonal RMnO$_3$ compounds provide text book examples to study multiferroicity.  Their crystal structure consists of triangular Mn planes packed along the $\it{c}$ axis and separated by layers of  rare earth ions (R= Ho-Lu) or non magnetic ions such as Y or In. As shown recently \cite{Fabreges2009}, the magnetic frustration does not arise only from the triangular geometry of antiferromagnetic (AF) first neighbour interactions in the $\it{ab}$ plane, but from competing interactions between Mn of adjacent planes.
In all compounds, the Mn moments order within a triangular plane in a three sublattice N\'eel structure, corresponding to 120$^\circ$ arrangements of the Mn moments in a triangle.
 Four possible AF structures can be stabilized, described by
  irreducible representations of the $P6_3cm$ space group with $\bf{k}=0$ propagation vector \cite{Munoz2000}.  These structures differ by the orientations of the Mn moments with respect to the $\it{a, b}$ crystal axes and by the relative orientations of Mn moments in adjacent planes. As shown in Ref. \onlinecite{Fabreges2009},
the selection of a given structure is controlled by the Mn position in the unit cell which depends on a unique parameter $\it{x}$ for the $\it{6c}$ sites.  The $\it{x}$ value with respect to a  critical threshold $\it{x_0}$=$1/3$ tunes the sign of the effective interaction between adjacent Mn planes. Within this frame, one can correlate the type of magnetic structure, the Mn position, and the sign of the effective exchange coupling in the compounds of the RMnO$_3$ family.

InMnO$_3$ is the only compound which does not fit simply with the above scheme. Actually, it corresponds to the peculiar situation where the Mn position is very close to $\it{x_0}$=$1/3$, so that interactions between adjacent Mn planes nearly cancel. Therefore one could expect new types of magnetic orders with two dimensional behavior or stabilized by further neighbour interactions. Moreover the InMnO$_3$ crystal structure has the smallest lattice constant $\it{a}$ and the largest lattice constant $\it{c}$ of the series \cite{Greedan1995}, so that in-plane and out-of-plane interactions differ much more than in the other compounds. The pioneering measurements of  Greedan {\em et al.} \cite{Greedan1995} showed that the magnetic structure of InMnO$_3$ indeed differs from those of the whole series, with a \textbf{k}=(0 0 $\frac{1}{2}$) propagation vector, corresponding to a doubled periodicity along $\it{c}$.  The sample showed broad magnetic reflections so that a two dimensional order was postulated.

 InMnO$_3$ is also interesting for its magnetoelectric properties. Ferroelectric hysteresis loop measurements performed on high quality samples showed no hysteresis below 250\,K, establishing that the pure compound is actually not ferroelectric \cite{Belik2009}, although ferroelectricity was earlier reported  \cite{Serrao2006} in some samples below 500\,K. In the pure samples, low frequency permittivity exhibits an anomaly near $T_{\rm N}$, showing evidence for a magnetoelectric coupling. Studies of Fe-substituted InMnO$_3$ showed that these compounds might constitute a new class of nearly room temperature multiferroics \cite{Belik2009b}. In InMnO$_3$, the electronic structure of the In$^{3+}$ ion with a fully filled 4d shell excludes the d$_0$-ness ferroelectricity at play in YMnO$_3$.  Considering the peculiar case of InMnO$_3$, a new covalent bonding mechanism was recently proposed to mediate ferroelectricity in hexagonal multiferroics \cite{Oak2011}.

  Since the measurements of Greedan {\em et al.}, no neutron study was made on InMnO$_3$. This could be due to the difficulty to synthesize big samples of high purity, and to the high absorption and low scattering power of the In$^{3+}$ ion which complicate the measurements.  To shed more light on the peculiar behavior of InMnO$_3$, we have synthesized powder sample of high purity in large amount under high pressure and high temperature conditions \cite{Belik2009}. We performed high resolution neutron study of the crystal structure versus temperature. We studied the magnetic order precisely by combining neutron diffraction and M\"ossbauer spectroscopy in a  $^{57}$Fe doped sample, and obtained the first results about the magnetic fluctuations.  We determine the magnetic structure precisely using group theory and we propose a possible explanation for its origin based on the influence of pseudo-dipolar interactions.

    \section{Experimental Details}
        Two samples were synthesized under high pressure.
         The first one is a stochiometric InMnO$_3$ sample of about $8\,g$ used for the neutron measurements. A second sample of $0.5\,g$
          with chemical formula InMn$_{0.99}$$^{57}$Fe$_{0.01}$O$_3$ was prepared for the M\"ossbauer measurements.
          For the synthesis, appropriate mixtures of In$_2$O$_3$ (99.9 \%purity) and Mn$_2$O$_3$ and Fe$_2$O$_3$  were placed in Au capsules and treated at $5\,GPa$ in a
         belt-type high pressure apparatus at 1500\,K for 90\,min (heating rate 120\,K/min).
         After the heat treatment, the samples were quenched to room temperature, and
         the pressure was slowly released. The resultant samples were dense black pellets. X-ray diffraction measurements showed that they contained a  small amount (1 mass \%) of cubic In$_2$O$_3$ impurity.

        The crystal structure and the evolution of the atomic parameter $\it{x}$ with temperature were determined by measuring a neutron powder diffraction (NPD) pattern at 300~K  and at selected temperatures on the high resolution powder diffractometer 3T2 of the Laboratoire L\'eon Brillouin (LLB) at  Orph\'ee reactor, with an incident neutron wavelength $\lambda=1.2253$~\AA. The magnetic structure was studied by collecting NPD patterns at several temperatures, between 200~K (above the magnetic transition) and 1.5~K. Both crystal and magnetic structures were refined using the Fullprof suite\cite{Rodriguez1993}. The $^{57}$Fe M\"ossbauer absorption spectra were recorded in the temperature range 4.2 - 140~K. We used a commercial $^{57}$Co:Rh$\gamma$-ray source, mounted on a triangular velocity electromagnetic drive.

    \section{Crystal Structure}
\begin{figure}[!h]
         	\centering
	        	 \includegraphics[width=8cm,height=5.5cm]{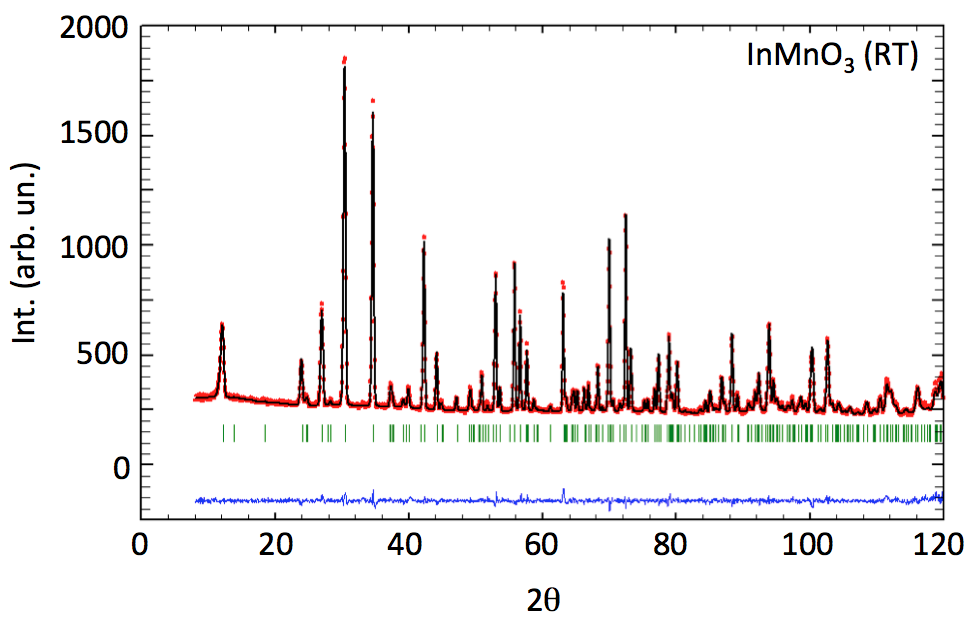}
               	\caption{(Color online) Observed and Fullprof calculated NPD
				pattern at room temperature. The Bragg reflections (tics), and the
				difference between the observed and calculated patterns are plotted at
				the bottom.}
               	\label{In_diff_RT}
		\end{figure}

        The refined NPD pattern at 300 K is shown in Fig.\ref{In_diff_RT}. All Bragg  reflexions of the pattern can be indexed within the hexagonal space group $P6_3cm$. The lattice constants $a=5.8837(1)$~\AA\ and $c=11.4829(1)$~\AA\ at 300 K  are in perfect agreement with previous results \cite{Greedan1995,Belik2009}. As noticed earlier, they strongly differ from those of the hexagonal RMnO$_3$ series, which scale from one compound to another \cite{Munoz2000,Munoz2001,Xu1995}.

		The refined atomic positions reported in Table \ref{position} agree with previous determinations from X-ray diffraction\cite{Greedan1995,Belik2009}. They are very close to those determined in compounds of similar ionic radius (R= Ho, Y, Yb). Each Mn atom is surrounded by oxygen ions forming a MnO$_5$ bipyramidal structure, with 3 O (2 O$_4$ and one O$_3$) ions close to the Mn plane, and two O (O$_1$ and O$_2$) ions at the apexes. Corner sharing MnO$_5$ bipyramids form layers separated along the \textit{c}-axis by In layers in which In ions occupy two distinct crystallographic sites (labelled $2a$ and $4b$).
		
        \begin{table}
        \centering
        	\begin{tabular}{|l|cccc|}
            	\hline
                Atoms & x & y & z & B$_{iso}$\\
                \hline
                \hline
                In(2a) & 0 & 0 & 0.274(2) & 0.845(120)\\
                In(4b) & $\frac{1}{3}$ & $\frac{2}{3}$ & 0.232(2) & 0.490(65)\\
                Mn(6c) & 0.345(4) & 0 & 0 & 0.334(43)\\
                O$_1$(6c) & 0.307(2) & 0 & 0.165(3) & 0.686(16)\\
                O$_2$(6c) & 0.640(1) & 0 & 0.336(3) & 0.686(16)\\
                O$_3$(4b) & 0 & 0 & 0.475(2) & 0.954(100)\\
                O$_4$(2a) & $\frac{1}{3}$ & $\frac{2}{3}$ & 0.020(2) & 0.575(54)\\
                \hline
                \hline
                Discrepancy & Bragg R-factor & 4.32\% & &\\
                Factors & RF-factor & 3.21\% & &\\
                \hline
            \end{tabular}
            \caption{Atom positions, thermal parameters and discrepancy factors at room temperature}
            \label{position}
        \end{table}

\begin{figure}
	       	\centering
	       	 \includegraphics[width=8cm]{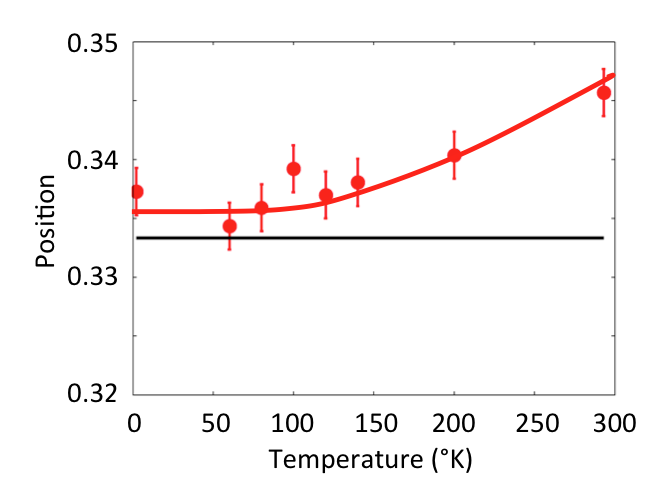}
	          	\caption{(Color online) Refined position $x$ of Mn versus temperature in
	           	reduced units of the cell parameter $a$. The horizontal black line is located at $x=1/3$,
	           	the red line is a guide to the eyes.}
	           	\label{In_pos_Mn}
		\end{figure}

\begin{figure}
	       	\centering
	       	 \includegraphics[width=8cm]{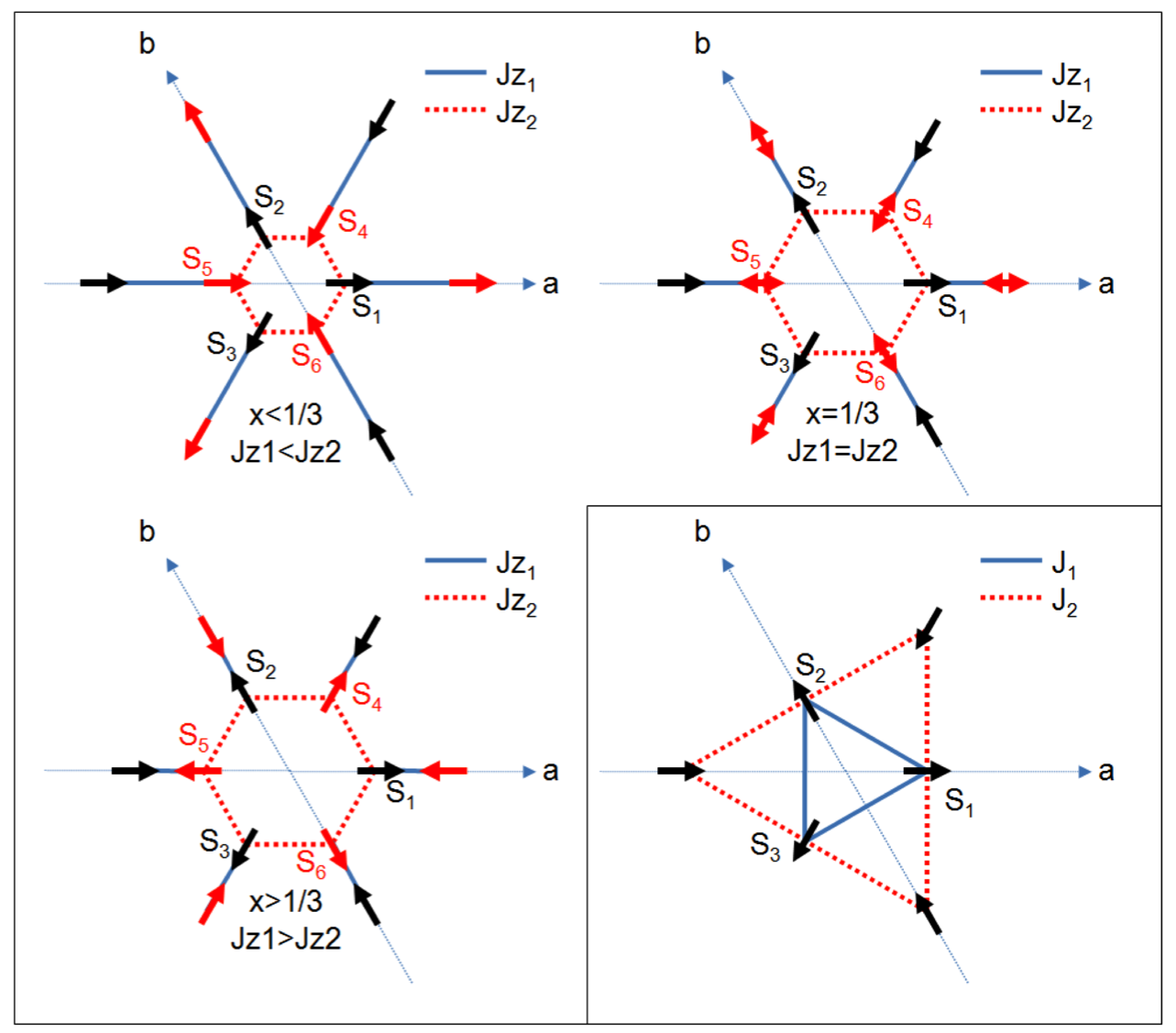}
	          	\caption{(Color online) Interplane exchange paths versus Mn position. Two exchange paths Jz$_1$ and Jz$_2$ are in competition and the $x=$1/3 Mn position corresponds to the specific case Jz$_1$ = Jz$_2$. Insert : inplane exchange paths leading to the 120$^{\circ}$ magnetic configuration.}
	           	\label{In_ech}
		\end{figure}

         The thermal variation of the positional parameter $x$ of the Mn sites is reported on fig. \ref{In_pos_Mn}. One notices  that $x$ decreases with decreasing temperature down to about 150\,K, then becomes very close to 1/3 in the 0$<$T$<$150\,K temperature range, which spans the whole ordered magnetic phase (T$_{\rm N}$=118\,K). Based on this sole observation it is possible to predict that the two possible interplane exchange paths between Mn ions are almost identical (Fig. \ref{In_ech}), which should dramatically decrease the effective exchange coupling along the $c$-axis.

	\section{Magnetic Structure} \label{strumag}
\begin{figure}
         	\centering
	        	 \includegraphics[width=8cm]{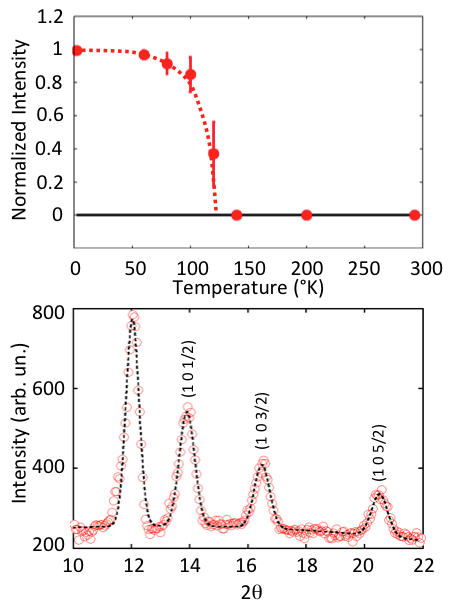}
               	\caption{(Color online) Upper panel : integrated intensity of the (1 0 1/2) Bragg reflection versus temperature. The red dashed line is a guide to the eyes. Lower panel : observed NPD pattern at low temperature (T=1.5\,K). The \textbf{k}=(0 0 $\frac{1}{2}$) propagation vector is easily observed through the existence of (1 0 $\frac{(l+1)}{2}$) Bragg reflections.}
               	\label{In_fp}
		\end{figure}

		The NPD pattern collected at T=1.5\,K on the high resolution diffractometer 3T2 is reported on Fig. \ref{In_fp}:bottom, focusing on the range in the scattering angle 2$\theta$ where magnetic Bragg reflections with half integer $\it{l}$ values can be observed. All magnetic peaks can be indexed within the hexagonal space group $P 6_3 cm$ with a propagation vector \textbf{k}=(0 0 0.50(1)). In contrast with the other members of the family, there is no magnetic contribution at the positions of the structural peaks. The (1 0 $\frac{2l+1}{2}$) Bragg reflections appear below $T_{\rm N}$=120(2)\,K, with a peak width  limited by the experimental resolution and their thermal variation is monotonic (Fig. \ref{In_fp}:top).  All these features shows the onset below $T_{\rm N}$=120(2)\,K of a three dimensional order for the Mn moments, with a magnetic unit cell doubled along the $\it{c}$ axis, and without spin reorientation transition below $T_{\rm N}$.

\begin{figure}
         	\centering
	        	\includegraphics[width=8cm]{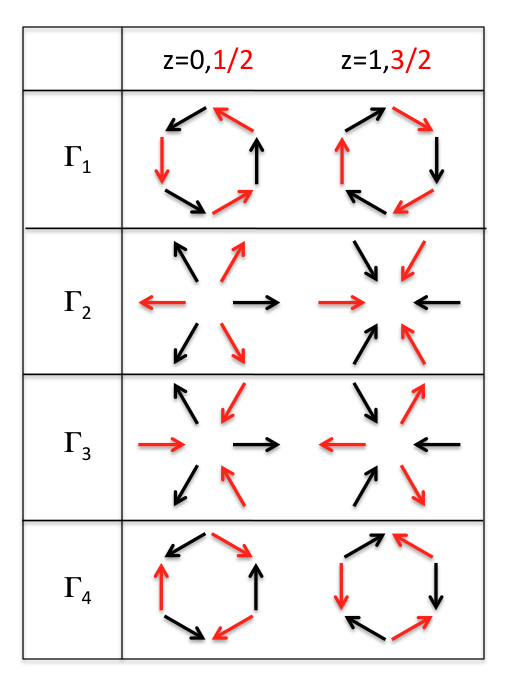}
               	\caption{(Color on line) Magnetic structures associated to the 4 unidimensional irreductible representations of the $P6_3cm$ space group. Red arrows indicates magnetic moments with real Fourier components, black arrows indicates moments with imaginary Fourier components.}
               	\label{In_Gamma}
		\end{figure}

        To analyze the magnetic structure we searched for all Irreductible Representations (IR)
        compatible with the crystal symmetry using the theory of group representation analysis\cite{Bertaut1963}
        and the program Basireps\cite{Rodriguez2001}. The atomic position of Mn ions in the unit cell was kept equal to (1/3 0 0)
        close to the position observed experimentally. In the space groupe $P6_3cm$, the $\it{6c}$ site of Mn ions allows 6
        irreductible representations labelled from $\Gamma_1$ to $\Gamma_6$ (Fig. \ref{In_Gamma}). The $\Gamma_1$ and $\Gamma_4$
        representations are defined by one basis vector associated with a 120$^{\circ}$ magnetic order within the $\it{ab}$ planes
        whereas the $\Gamma_2$ and $\Gamma_3$ are defined by two basis vectors, the second one
        allowing an out-of-plane component. The $\Gamma_5$ and $\Gamma_6$ representations
        correspond to magnetic orders with unequivalent  magnetic moments on each sites which have not been
        considered, as for the rest of the RMnO$_3$ family\cite{Munoz2000}.
        %and several studies of magnetic interactions in such systems\cite{ref,ref,ref} tend to discredit these two last solutions.

        The Fourier component corresponding to the propagation vector $\bf{k}$ for a Mn site $\it{n}$ of the unit cell is expressed as : $M_n(z)=M\,e^{-i\textbf{k}.\textbf{r}_n}$ where $r_n$ denotes the position of the $n^{th}$ Mn ion in the unit cell, referred by its $\it{z}$ coordinate along the $\it{c}$ axis.
         In our particular case, the $\bf{k=}$(0 0 $\frac{1}{2}$) propagation vector yields a purely real Fourier component of the magnetic moment in the z=0,1,2,.. Mn planes and purely imaginary components in the z=1/2,3/2,.. planes. In order to overcome this difficulty and to be consistent with the presence of equivalent moments on all Mn sites deduced from the M\"ossbauer results (see below), we have introduced a global phase shift $\phi=2~\pi/8$ in the expression of the structure factor. The phase and amplitude of the Fourier components were used to determine the magnitude of the ordered moment at a given Mn site.

         %Therefore, half the planes remain paramagnetic which is in conflict with our M\"ossbauer data showing that each Mn sites, share the same magnetic hyperfine field (see below). The M\"ossbauer results clearly indicate that there is only one magnetic environment on the Mn sites, raising equivalent magnetic ordered moments.

        %To overcome this incompatibility, one has to remind that the atomic positions within the unit cell are defined relatively to each others, which means that a shift of the base pattern is allowed without changing the properties of the crystal. In other words, a phase $\phi$ can be freely adjust within the exponential and choosing leads to equivalent magnetic moments on each planes. It is however important to notice that the Fullprof refinment is then made on the amplitude of the magnetization $M$ that has to be corrected by a $cos(\pi/4)=\sqrt{2}/2$ factor in order to reflect the numerical value of ordered magnetic moment.

		As for the rest of the RMnO$_3$ family, we find that magnetic configurations associated to $\Gamma_1$ and $\Gamma_3$ IR are homometric (namely they share the same structure factor) so they cannot be distinguished in a powder neutron diffraction experiment. The same holds for the $\Gamma_2$ and $\Gamma_4$ magnetic configurations. Our refinements yield a discrepancy factor $R_{mag}=12.54\,\%$ for the $\Gamma_2$ and $\Gamma_4$ IR, much better than for $\Gamma_1$ and $\Gamma_3$ ($R_{mag}=19.8\,\%$). The $R_{Bragg}$ factor in the ordered magnetic phase was close to $5\,\%$.  The best fit of our data was obtained for an ordered magnetic moment of $3.25\,\mu_B$ at 1.5\,K, very similar to the moment found in the rest of the hexagonal RMnO$_3$ family\cite{Munoz2000}. We conclude that the Mn moments order in the $\it{a,b}$ planes, in bilayers  ordered according to either a $\Gamma_2$ or a $\Gamma_4$ configuration, as for YbMnO$_3$ or ScMnO$_3$ with \textbf{k}=\textbf{0} propagation vector, but with antiferromagnetic relative orientations of two neighboring bilayers.

    \section{$^{57}$Fe M\"ossbauer data}
 		Three $^{57}$Fe M\"ossbauer spectra were recorded, at T=140, 80 and 4.2\,K. The
		spectra at 4.2 and 140\,K are represented in Fig.5. At 140\,K, a quadrupolar
		hyperfine spectrum is observed, with a quadrupolar splitting $\vert \Delta E_Q
		\vert$=0.5(1)\,mm/s, typical for Fe$^{3+}$ in the paramagnetic phase. Below $T_{\rm N}$, at
		4.2 and 80\,K, a six-line spectrum is observed, attributable to a single
		magnetic hyperfine field, with a small quadrupolar shift $\epsilon$=
		0.26(1)\,mm/s. This indicates that all the $^{57}$Fe nuclei experience the same
		hyperfine field (48.6\,T at 4.2\,K and 43\,T at 80\,K), hence all the
		substituted Fe ions bear the same magnetic moment. One can conclude that the
		ordered magnetic moment of the Mn ion is the same on each site.

 		\begin{figure}
         	\centering
	        	 \includegraphics[width=8cm]{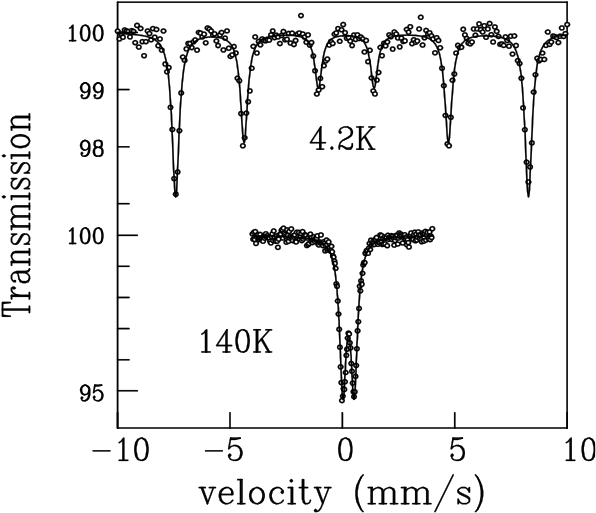}
               	\caption{$^{57}$Fe M\"ossbauer spectra in InMn$_{0.99}$Fe$_{0.01}$O$_3$
               	below and above $T_N$=120\,K. At $T=140\,K$ a quadrupolar doublet characteristic of paramagnetic Fe$^{3+}$ is
               	observed. At $T=4.2\,K$ the spectrum shows a six-lines hyperfine pattern
               	perfectly reproduced by a single hyperfine magnetic field.}
               	\label{In_moss}
		\end{figure}
		
		It is possible to obtain information about the angle $\theta$ between the
		hyperfine field and the principal axis of the electric field gradient (EFG)
		tensor, responsible for the quadrupolar hyperfine interaction. Indeed, the
		relationship between the quadrupolar splitting obtained in the paramagnetic
		phase and the quadrupolar shift measured in the magnetically ordered phase is:
		$\epsilon = \Delta E_Q\  \frac{3 \cos^2\theta - 1}{2}$. Since the sign of
		$\Delta E_Q$ cannot be determined, one derives two acceptable values for $\theta$:
		90$^\circ$ and 35.3$^\circ$. The local symmetry of the Fe(Mn) sites is {\it 6c},
		which implies that the EFG tensor has one axis along {\bf c} and the two other
		axes in the \textit{a,b} plane, but the principal axis cannot be determined only by
		symmetry considerations. Assuming it lies along {\bf c}, then the solution
		$\theta=90^\circ$ would be adequate, in analogy with the rest of the RMnO$_3$
		family.

	\section{Short range correlations in the paramagnetic phase}
\begin{figure}
         	\centering
	        	 \includegraphics[width=8cm]{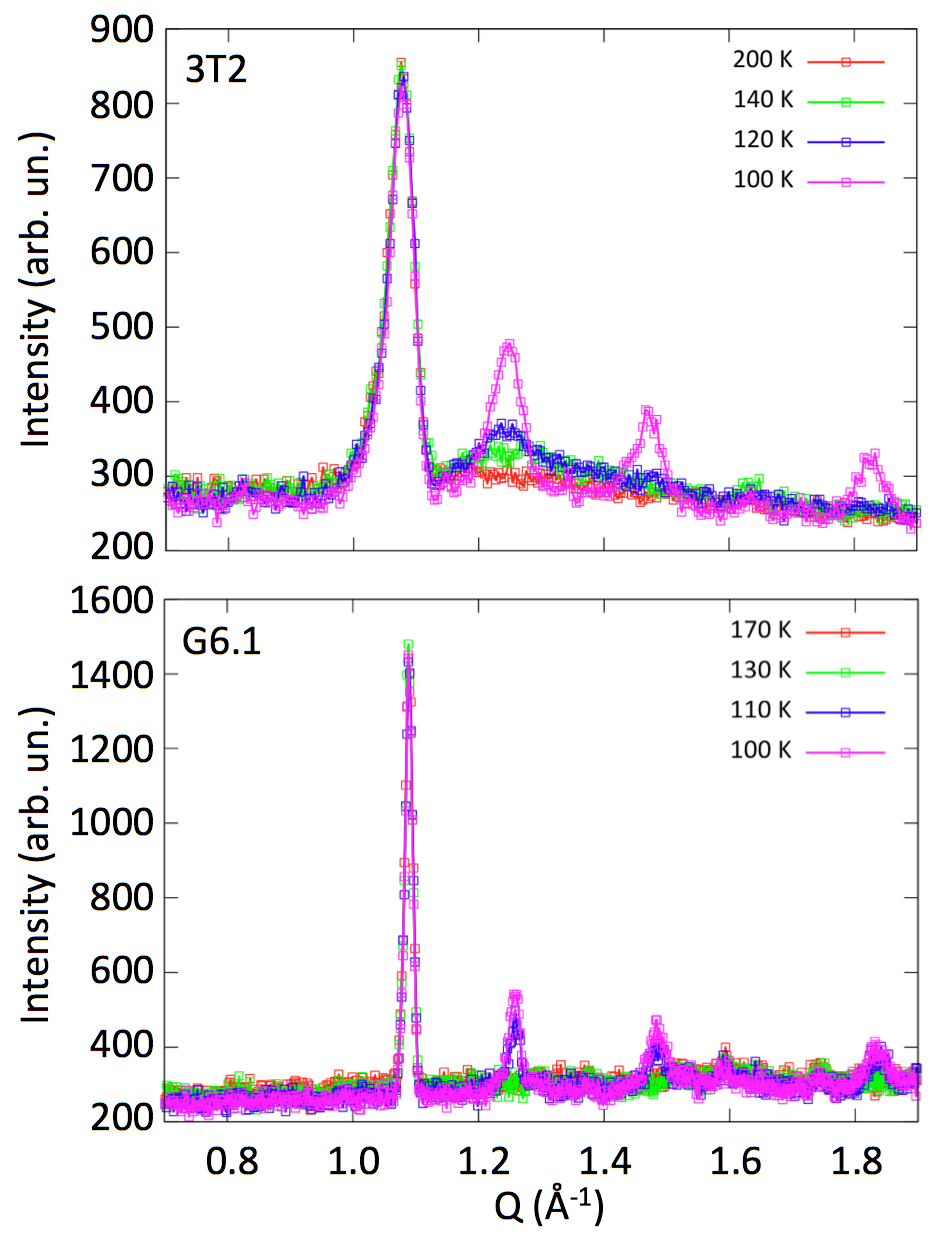}
               	\caption{(Color online) Observed and Fullprof calculated NPD
				patterns at several temperatures.Above T$_{\rm N}$ a strong diffuse scattering is observed on the patterns recorded
				on 3T2 spectrometer (top) with $\lambda=1.225\,\AA$. This scattering is not visible on the G6.1 patterns (bottom) for which $\lambda=4.74\,\AA$.}
               	\label{In_diffus}
		\end{figure}

\begin{figure}
         	\centering
	        	 \includegraphics[width=8cm]{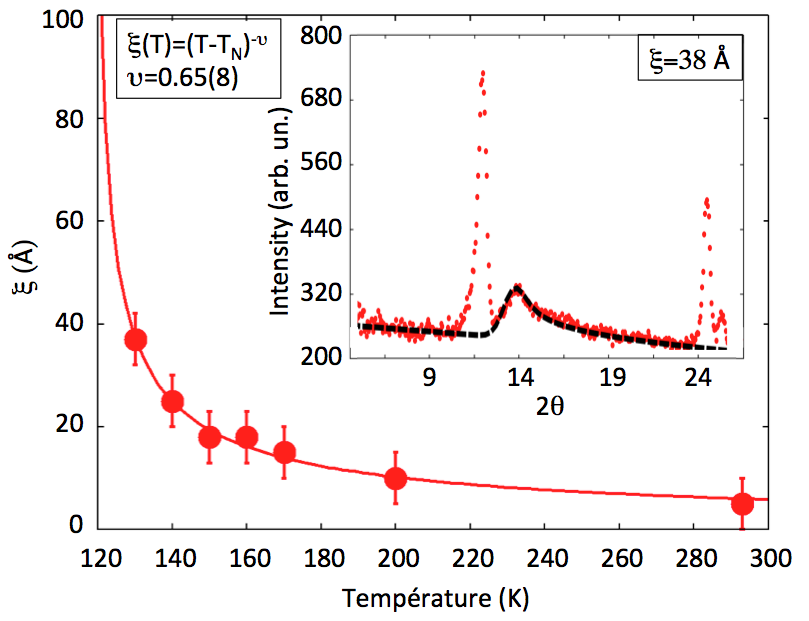}
\caption{(Color online)Refined correlation length versus temperature (red dots) and fit of the critical exponent $\nu$ (solid line). Insert: intensity recorded at T=130\,K on the 3T2 spectrometer. The dashed line is a fit of the diffuse intensity with a Warren function.}
               	\label{In_xi}
		\end{figure}

       	The powder diffraction patterns measured on 3T2 above $T_{\rm N}$ (Fig. \ref{In_diffus}) show a strong diffuse scattering, already observed by Greedan \textit{et al} \cite{Greedan1995}. The asymmetric shape of this scattering is directly connected with the presence of two dimensional correlations between Mn moments of a given plane. Using a Warren-like profile \cite{Warren1941} we refined the lengthscale $\xi$ of these correlations  (Fig. \ref{In_xi}). The $\xi$ values above $T_{\rm N}$ agree with those deduced previously \cite{Greedan1995}. However in the sample studied in Ref.\onlinecite{Greedan1995}, the 2D correlations persist below $T_{\rm N}$, coexisting with  half integer Bragg reflections of finite width, whereas in the present case $\xi$ diverges at $T_{\rm N}$, showing the onset of a purely three dimensional long range magnetic order.
       	
       Interestingly, spectra collected in the same temperature range on the G6.1 diffractometer using a large incident neutron wavelength showed no signature of this diffuse scattering (Fig. \ref{In_diffus} bottom). To understand this peculiarity, one should notice that a neutron diffractometer probes both elastic and inelastic signal and integrates all contributions at a given scattering angle. The energy range over which this integration is performed depends on the energy of the incident neutron.  Knowing that G6.1 is a cold diffractometer with an incident energy $\hbar^2k_i^2=4\,meV$ ($\lambda=4.74\,\AA$) and 3T2 a thermal one with $\hbar^2k_i^2 \approx$ 40\,meV ($\lambda=1.225\,\AA$), one concludes that the observed diffuse scattering above $T_{\rm N}$ corresponds to dynamical short range correlations between Mn moments, involving high energy fluctuations, at a scale of tens of meV.
       	
		 The analysis of the paramagnetic scattering suggests a picture of uncorrelated Mn planes, in which dynamical magnetic correlations develop with decreasing temperature down to $T_{\rm N}$. The 3D magnetic ordering stabilized at $T_{\rm N}$ should be triggered by a weak interaction between Mn moments belonging to different planes, whose origin is discussed below.

    \section{Discussion} \label{disc}
        \subsection{Magnetic ordering and frustration}
         We first recall the scheme of interactions used in Ref. \onlinecite{Fabreges2009} to discuss the magnetic structures observed in the hexagonal RMnO$_3$ family with $\bf{k}$=$\bf{0}$ propagation vector. In these compounds, a given magnetic structure of symmetry  $\Gamma_i$ (i=1-4) is stabilized by near neighbour exchange interactions as well as planar and uniaxial anisotropies, so that the Hamiltonian of the system is composed of three terms :
        	
        	\begin{eqnarray}
        		\mathcal{H}_{Heis} & = & \sum_{i,j} J_{ij}\, \textbf{S}_i.\textbf{S}_j + \sum_i D\,(S_i^z)^2 - \sum_i \textbf{h}_i.\textbf{S}_i
        	\end{eqnarray}
where $S_i$ is the Mn spin on the $i^{th}$ site, J$_{ij}$ the exchange constants, $D$ a planar anisotropy and $\textbf{h}_i$ a local field yielding a preferential orientation for the $\textbf{S}_i$ spin.

		The exchange term has two distinct parts, involving in-plane and out-of-plane interactions respectively. Due to the triangular lattice, the in-plane interactions yield a two dimensional 120$^{\circ}$ order, with no preferential orientation of the magnetic moments with respect to the crystal axes. Out-of-plane interactions couple Mn moments from adjacent planes yielding the 3D order. In this scenario, the Mn position is crucial since two possible exchange paths compete along the $\it{c}$ axis. The selection of a given structure is controlled by the Mn position. In InMnO$_3$, the Mn position is close to the critical threshold of 1/3 for which the two exchange paths are strictly equal. This leads to a full compensation of the exchange interactions along the $\it{c}$ axis and to an effective out-of -plane exchange interaction close to zero. This specific position of the Mn ions could explain the dynamical short range 2D order observed above T$_{\rm N}$, and attributed to uncorrelated Mn planes.
		
		The two other terms of equation (1) are respectively the planar anisotropy $D$ which confines the Mn moments in the basal plane, and the local field $\textbf{h}$ which plays the role of a uniaxial anisotropy and selects preferential directions either along or perpendicular to the crystal axes.

		These terms however cannot explain the long period 3D structure with $\bf{k}$=(0 0 $\frac{1}{2}$) stabilized in InMnO$_3$. Therefore one needs to consider further neighbor interactions, with different symmetries than the exchange interactions, such as the Dzyaloshinskii-Moriya (DM) or the pseudo-dipolar interaction \cite{vanVleck1937}.  A similar approach \cite{Fabreges2008} was proposed to account for the ordering of the Yb moments in YbMnO$_3$. In the following, we focus on the pseudo-dipolar interaction since the DM interaction is hardly compatible with long exchange path (Mn-O-O-Mn and Mn-O-O-O-O-Mn) between Mn of different planes. The pseudo-dipolar interactions is written as :
			
        	\begin{eqnarray}
        		\mathcal{H}_{dip} & = & - \sum_{i,j} \textbf{S}_i\,J_{ij}^{dip}\,\textbf{S}_j \nonumber  \\
& = &
-\alpha~\sum_{i}\sum_{j}
\left[
3 \frac{(\textbf{S}_j.\textbf{r}_{ij}).\textbf{r}_{ij}}{r_{ij}^2} - \textbf{S}_j
\right] \textbf{S}_i
		\end{eqnarray}
where $\alpha$ is a constant and $\textbf{r}_{ij}$ joins sites $i$ and $j$. The matricial representation of the
pseudo dipolar interaction $J_{ij}^{dip}$ coupling two different Mn sites reads as :
			
			\begin{eqnarray}
				J_{ij}^{dip} & = & \alpha
\left[\frac{3}{r_{ij}^2}~\left( \begin{array}{c c c} r_{ij}^xr_{ij}^x & r_{ij}^xr_{ij}^y & r_{ij}^xr_{ij}^z \\ r_{ij}^yr_{ij}^x & r_{ij}^yr_{ij}^y & r_{ij}^yr_{ij}^z \\ r_{ij}^zr_{ij}^x & r_{ij}^zr_{ij}^y & r_{ij}^zr_{ij}^z \end{array} \right) - \,\mbox{l\hspace{-0.50em}1} \right]
			\end{eqnarray}
			
where $\,\mbox{l\hspace{-0.50em}1}$ is the identity matrix. Assuming the $\bf{k}$=(0 0 $\frac{1}{2}$) magnetic structure described above, we calculate the magnetic field $\textbf{B}_i$ induced on the $i^{th}$ site by the surrounding Mn at sites $j$, $\textbf{B}_i = \sum_j J_{ij}^{dip} \textbf{S}_j$. First, we find that the contribution arising from the neighbouring sites in adjacent $z=\pm1/2$ planes is zero. Thus, there is no pseudo dipolar coupling between adjacent layers in agreement with the idea of purely two dimensionnal dynamical correlations above $T_{\rm N}$. In contrast, the contribution from sites in $z=\pm 1$ planes is different from zero. Moreover, the classical energy calculated as $E = - \textbf{B}_i \textbf{S}_i$ is negative (assuming $\alpha$ is positive). In other words, the pseudo-dipolar interaction stabilizes the 3d magnetic structure observed in InMnO$_3$ and drives the $\bf{k}$=(0 0 $\frac{1}{2}$) propagation vector. 			
				
		\subsection{Spin wave spectrum }
To confirm the possible role of the pseudo-dipolar coupling, we propose to carry out spin dynamics measurements, as specific features associated to the pseudo-dipolar coupling should be easily seen on spin wave dispersion relations. This issue could be sorted out by inelastic neutron scattering experiments performed on a triple axis spectrometer.

From the interaction scheme described above, one can calculate the  spectrum of the spin wave excitations in the ordered phase. We use the previous Heisenberg Hamiltonian, to which we add the pseudo dipolar term, written as :
			
			\begin{eqnarray}
        		\mathcal{H} & = & \mathcal{H}_{Heis} - \sum \textbf{S}_i\,J_{ij}^{dip}\,\textbf{S}_j
        	\end{eqnarray}
			
			Each term affects the spin wave spectrum in a specific way. The Heisenberg Hamiltonian $\mathcal{H}_{Heis}$ is responsible for the magnitude of the dispersion, namely the in-plane exchange interaction induces the dispersion along the (q$_h$ 0 0) and (0 q$_k$ 0) directions of the reciprocal space, whereas the out-of-plane exchange yields the dispersion along the (0 0 q$_l$) direction. Considering that the exchange interactions along $\it{c}$ nearly cancel due to the specific Mn position, one can predict that no dispersion should be observed along the (0 0 q$_l$) direction, yielding two flat modes. The anisotropy terms induce gaps in the dispersion curves. In RMnO$_3$, the planar anisotropy term induces a large gap of about 6 meV\cite{Petit2007} and the uniaxial term a smaller one, strongly dependent on temperature and likely enhanced by interaction with the rare earth moment \cite{Fabreges2011b}.
			
As concerns the influence of the pseudo dipolar term on the spin wave spectrum, one notices that this term involves both diagonal and off diagonal elements introducing new coupling between spin components. The diagonal elements act mainly as a combination of exchange and uniaxial anisotropy. Its effect should be easily seen at the zone center, the uniaxial gap increasing with the dipolar interaction strength $\alpha$.

\begin{figure}
	         	\centering
		        	\includegraphics[width=\linewidth]{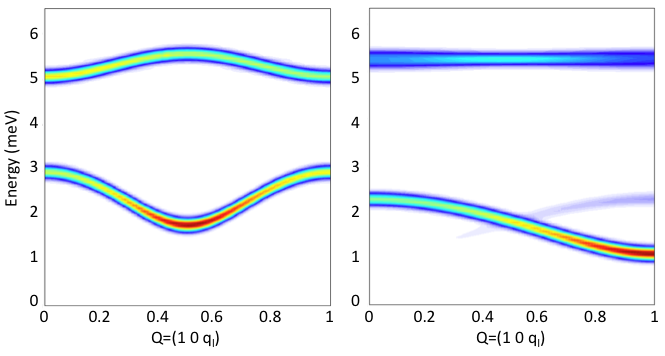}
	               	\caption{(Color online) Left : numerical calculation of the dynamical structure factor of spinwaves along the (0 0 q$_l$) direction in case of pseudo-dipolar coupling between Mn. Right : numerical calculation of the dynamical structure factor of spinwaves along the (0 0 q$_l$) direction in case of antiferromagnetic interplane exchange coupling between Mn.}
	               	\label{In_SW}
			\end{figure}

To illustrate this point, spinwave calculations of the dynamical structure factor were made with the following parameters : $J$=2.6 meV, $D$=0.55 meV and $h$=0.1 meV in the case of the magnetic structure of InMnO$_3$ refined above. The results along the (0 0 $q_l$) direction of the reciprocal space are reported on Fig. \ref{In_SW} in case of pseudo-dipolar (left) and interplane exchange (right) coupling. The coupling constant $\alpha$ and $J_{inter}$ were taken equal to 0.01\,meV (antiferromagnetic). In both cases, the spinwave dispersion curves are characterized by two gaps around 5\,meV and 2\,meV induced respectively by $D$, and $h$.

Considering the shape of the dispersion curves, the pseudo-dipolar interaction induces a dispersion of both the 2\,meV and 5\,meV modes. A maximum (respectively minimum) is observed at $\bf{Q}$=(1 0 0) and a minimum (respectively maximum) is observed at $\bf{Q}$=(1 0 $\frac{1}{2}$). On the other hand, the interplane exchange induces a dispersion of the low energy mode with a maximum at $\bf{Q}$=(1 0 0) and a minimum at $\bf{Q}$=(1 0 1), whereas the 5\,meV mode remains almost flat. The pseudo-dipolar interaction is at the origin of a change in the periodicity of the dispersion in perfect agreement with the $\textbf{k}=$(0 0 $\frac{1}{2}$) propagation vector.

Inelastic neutron scattering is mandatory to confirm the scheme of interaction proposed here for InMnO$_3$ as both behaviors are easily distinguishable and should be seen on a triple axis or time-of-flight spectrometer.  Up to now, precise measurements were hampered by the low intensity given by the available samples and by the powder averaging, but we hope to perform them in future.

    \section{Conclusion}
        In conclusion, our experimental study of InMnO$_3$  by neutron powder neutron diffraction and M\"ossbauer spectroscopy shows the onset of a  three dimensional magnetic order below $T_{\rm N}$= 120(2)K. The magnetic order with $\bf{k}$=(0 0 $\frac{1}{2}$) propagation vector shows a doubling of the magnetic unit cell along the $\it{c}$ axis, in contrast with the other compounds of the RMnO$_3$ family. This feature is directly related to the peculiar value of the Mn positional parameter in InMnO$_3$, close to the $1/3$ threshold where the effective exchange interaction along the $\it{c}$ axis cancels. We suggest that weak out-of-plane  pseudo-dipolar Mn interactions are responsible for the long period of the magnetic order. This weak coupling together with the  strong in-plane coupling yields the onset of two dimensional correlations between fluctuating moments, which settle above $T_{\rm N}$. InMnO$_3$ provides an original example of the links between magnetic frustration and multiferroicity, which should be further studied by inelastic neutron scattering.

        This work was partially supported by World Premier International Research Center (WPI) Initiative on Materials Nanoarchitectonics (MEXT, Japan), by the Japan Society for the Promotion of Science (JSPS) through its Funding Program for World-Leading Innovative R\&D on Science and Technology (FIRST Program), and by the Grants-in-Aid for Scientific Research (22246083) from JSPS, Japan.

%%%%%%%%%%%%%%%%%%%%%%%%%%%%%%%%%%%%%%%%%%%%


\begin{thebibliography}{}
% intro
\bibitem{Katsura2005} H. Katsura, N. Nagaosa, and A. V. Balatsky, Physical Review Letters {\bf 95}, 057205 (2005).
\bibitem{Sergienko2006} I. A. Sergienko and E. Dagotto, Physical Review B {\bf 73}, 094434 (2006).
\bibitem{Fabreges2009} X. Fabreges, S. Petit, I. Mirebeau, S. Pailhes, L. Pinsard, A. Forget, M. T. Fernandez-Diaz, and F. Porcher, Physical Review Letters {\bf 103}, 067204 (2009).
\bibitem{Munoz2000} A. Munoz, J. A. Alonso, M. J. Martinez-Lope, M. T. Casais, J. L. Martinez, and M. T. Fernandez-Diaz, Physical Review B {\bf 62}, 9498 (2000).
\bibitem{Greedan1995} J. E. Greedan, M. Bieringer, J. F. Britten, D. M. Giaquinta, and H. C. Zurloye, Journal of Solid State Chemistry {\bf 116}, 118 (1995).
\bibitem{Belik2009} A. A. Belik, S. Kamba, M. Savinov, D. Nuzhnyy, M. Tachibana, E. Takayama-Muromachi, and V. Goian, Physical Review B {\bf 79}, 054411 (2009).
\bibitem{Serrao2006} C. R. Serrao, S. B. Krupanidhi, J. Bhattacharjee, U. V. Waghmare, A. K. Kundu, and C. N. R. Rao, Journal of Applied Physics {\bf 100} (2006).
\bibitem{Belik2009b} A. A. Belik, T. Furubayashi, Y. Matsushita, M. Tanaka, S. Hishita, and E. Takayama-Muromachi, Angewandte Chemie-International Edition  {\bf 48}, 6117 (2009).
\bibitem{Oak2011}M. A. Oak, J. H. Lee, H. M. Jang, J. S. Goh, H. J. Choi, and J. F. Scott, Physical Review Letters, {\bf 106}, 047601 (2011).
\bibitem{Rodriguez1993} J. Rodriguez-Carvajal, Physica B {\bf 192}, 55 (1993).
\bibitem{Munoz2001} A. Munoz, J. A. Alonso, M. J. Martinez-Lope, M. T. Casais, J. L. Martinez, and M. T. Fernandez-Diaz, Chemistry of Materials {\bf 13} 1497-1505 (2001).
\bibitem{Xu1995} H. W. Xu, J. Iwasaki, T. Shimizu, H. Satoh, and N. Kamegashira, Jopurnal of Alloys and Compounds {\bf 221}, 274-279 (1995).
\bibitem{Bertaut1963} E. F. Bertaut and M. Mercier, Physics Letters, {\bf 5}, 27 (1963).
\bibitem {Rodriguez2001} J. Rodriguez-Carvajal, \url{http://www.ill.eu/sites/fullprof/php/programsfa7c.html?pagina=GBasireps}.
\bibitem{Warren1941} B. E. Warren, Physical Review, {\bf 59}, 693 (1941).
\bibitem{vanVleck1937} J. H. van Vleck, Physical Review, {\bf 52}, 1178 (1937).
\bibitem{Fabreges2008} X. Fabreges, I. Mirebeau, P. Bonville, S. Petit, G. Lebras-Jasmin, A. Forget, G. Andre, and S. Pailhes, Physical Review B, {\bf 78}, 214422 (2008).
\bibitem{Petit2007} S. Petit, F. Moussa, M. Hennion, S. Pailhes, L. Pinsard-Gaudart, and A. Ivanov, Physical Review Letters {\bf 99}, 266604 (2007).
\bibitem{Fabreges2011b} X. Fabreges, S. Petit, and I. Mirebeau, To be published (2011)
\end{thebibliography}
\end{document}